\documentstyle[12pt]{article}
\def\({[}
\def\){]}
\begin{document}
\begin{flushright}
SU-ITP-96-28\\
hep-th/9606423\\
\today\\
\end{flushright}
\vspace{1cm}
\begin{center}
\baselineskip=16pt

{\Large\bf    Hybrid Inflation from Supergravity D--Terms}  
\vskip 1.5 cm

{\bf Edi Halyo${}^{a}$\footnote{E-mail address:
 halyo@dormouse.stanford.edu}~}
\vskip 1 cm
${}^{a}$Department of Physics\\ 
Stanford University\\
Stanford CA 94305\\

\vskip 1.5cm
\end{center}
\begin{center}
{\bf Abstract}
\end{center}

We argue that the mass of the inflaton can be much smaller than the Hubble
constant in supergravity models in which inflation is driven by D--terms and
not F--terms. We investigate a supergravity toy model  which leads to hybrid 
inflation 
due to an anomalous D--term. We show that the slow--roll condition 
can be satisfied and the correct magnitude for density perturbations can be 
obtained for some choice of model parameters. The kind of model considered 
can naturally arise in some string derived supergravity models.

\newpage

\baselineskip=18pt
\section{Introduction}

It is well--known that the main difficulty in having inflation in supergravity 
is the problem associated with the inflaton mass\cite{STW}. For inflation to happen, 
the inflaton mass
has to be much smaller than the Hubble constant,$H$, which is given by the vacuum energy 
during inflation by $V_0=3H^2 M_P^2$\cite{AL}. (Here $M_P=M_{Planck}/ \sqrt{8\pi}$ is the reduced Planck mass.) On the other hand, in supergravity models where the positive vacuum energy required for inflation arises
due to nonzero F--terms, the inflaton mass (as all other scalar masses) is about the Hubble
constant\cite{DIN}. Thus the slow--roll condition is not satisfied and inflation cannot occur.
This is a general result and does not depend whether the inflaton Kahler potential is minimal or not. The origin of this problem lies in the specific form of the scalar potential, namely
the $e^K$ factor which multiplies the F--terms.

A natural and elegant solution to the inflaton mass problem is to consider inflation which is driven by supergravity D--terms associated with $U(1)$ gauge groups
rather than the F--terms\cite{CAS}.
\footnote{While this paper was in preparation, Ref. [5] which is
along the same lines appeared.} 
If the vacuum energy during inflation
is dominated by the D--terms (whether F--terms vanish or not) the inflaton mass can be much smaller
than the Hubble constant. In particular, if the inflaton is a gauge singlet and therefore does
not appear in the D--term it does not obtain a mass at tree level (during inflation) but only at the one loop level due to supersymmetry breaking effects. The one--loop inflaton mass is 
always smaller than the Hubble constant thus leading to inflation naturally. 

Below we consider a hybrid inflation\cite{HI} scenario which is driven by an anomalous D--term which arises due to an anomalous $U(1)$ gauge group
in supergravity. Anamalous D--terms are generic in certain classes of superstring derived supergravity models. We will see that it is the anomaly contribution to the D--term which 
gives rise to the vacuum energy and drives inflation. In
addition, there is a simple superpotential which gives F--terms but these vanish during and after inflation. The F--term contribution to the potential is nevertheless essential since it gives
rise to the interaction of the two fields required for hybrid inflation. We find that the slow--roll
condition can be satisfied together with the constraint from $\delta \rho /\rho$ for some
choice of parameters. 

The letter is organized as follows. In section 2, we describe our supergravity toy model and show that it leads to hybrid inflation.
In section 3, we show that this kind of models can naturally arise in realistic superstring models.
Section 4 contains a discussion of our results and conclusions.

\section{Inflation Driven by Supergravity D--terms}

In supergravity the scalar potential is given by\cite{SUG}
\begin{equation}
V=e^{K/M_P^2} [F_iF_j (G_{i}^j)^{-1}-3{|W|^2 \over M_P^2}]+g^2Ref^{-1} |D_a|^2
\end{equation}
where the first and second terms are the contributions due to the F and D--terms respectively. $g \sim 1/2$ is the $U(1)$ coupling constant. The
F--term is given by
\begin{equation}
F_i=W_i+K_iW
\end{equation}
and the anomalous D--term is
\begin{equation}
D_a=\phi_i K_i+M^2
\end{equation}
The anomaly contribution $M^2$ can be calculated in string models and is given by\cite{DSW}
\begin{equation}
M^2={TrQ_A \over 192 \pi^2} g^2 M_P^2
\end{equation}
Here $Q_A$ is the charge of the fields under the anomalous $U(1)$ group and trace indicates
a sum over all fields.
Above,  $K$, and $W$ are the Kahler potential and the superpotential which fix the
supergravity model (in addition to the gauge function which we took to be minimal) respectively. 

We see that for inflation to occur either an F--term or a D--term must be nonzero since inflation
requires a positive vacuum energy, $V_0>0$.
If inflation is driven by a nonzero F--term (and the D--term vanishes) then the inflaton mass
(as all other scalar masses) is given by\cite{STW,DIN}
\begin{equation}
m^2_{\phi} \sim K_{\phi \bar \phi} V+...
\end{equation}
The right hand side is evaluated at the minimum of $V$ and the dots indicate the other terms
which arise from differentiation of the square brackets in Eq. () which are of the same order
of magnitude. Using the realtion
$V_0=3H^2 M_P^2$ we find that the inflaton mass is about the Hubble constant, $m^2_{\sigma} \sim H^2$.
This is true whether the inflaton Kahler potential is minimal ($K=\phi \bar \phi$) or
logarithmic as in string derived models ($K=-log(\phi +\bar \phi$).
Thus, inflation cannot occur in these kinds of models since the slow--roll condition
$m^2_{\phi}<<H^2$ is not satisfied.
The origin of the inflaton mass problem is the $e^K$ factor in the supergravity scalar potential.

A natural solution is to consider inflation driven by a D--term and not by the F--term. Then,
the factor $e^K$ is automatically absent from the scalar potential during and after inflation since $F=0$. From Eqs. (1) and (3) we see that if the inflaton
does not appear in the anomalous D--term it does not obtain a mass at tree level. 
In any case, the inflaton has to be a gauge singlet lest it obtains a large mass from gauge boson loops. A gauge singlet inflaton gets a small mass from one--loop effects which arise due to
supersymmetry breaking.

We now consider our supergravity toy model which exhibits the essential features
required for hybrid inflation\cite{HI} driven by D--terms. In the next section we show that it
has the generic properties of some string derived supergravity models. Consider a
model with two fields $\sigma$ and $\phi$ with the superpotential
\begin{equation}
W=c \sigma \phi^2
\end{equation}
which can be imposed by discrete symmetries. $c$ is a constant of $O(1)$. The Kahler potential we consider is
\begin{equation}
K=-log(\sigma + \bar \sigma)+\phi \bar \phi
\end{equation}
Note that we take the Kahler potential for $\phi$ to be minimal and that of $\sigma$ to be nonminimal. In the next section we justify this by assuming that $\phi$ is a twisted matter
field and $\sigma$ is a modulus--like fields from the untwisted sector of the string spectrum.
Including the anomalous D-term contribution, the scalar potential becomes

\begin{eqnarray}
V={e^{\phi \bar \phi/M_P^2} |c|^2 \over (\sigma+ \bar \sigma)}[|\phi|^2 |\sigma|^2 |2+|\phi|^2|^2
+ |\sigma+ \bar \sigma|^2|\phi|^4 \cr
|1-|{\sigma \over (\sigma+ \bar \sigma)}|^2-3|\sigma|^2 |\phi|^4]
+g^2|-|\phi|^2+M^2|^2
\end{eqnarray}

(Here we omited negative powers of $M_P$ since those terms are negligible for $\phi<<M_P$ which holds during hybrid inflation.)
Taking both $\sigma$ and $\phi$ to be real and canonically normalizing $\sigma$ by the
transformation $\sigma \rightarrow \sigma(\sigma+ \bar \sigma)$ we obtain
\begin{equation}
V=e^{\phi \bar \phi} |c|^2 \sigma^2 \phi^2 (4+2\phi^2+\phi^4)+g^2|-\phi^2+M^2|^2
\end{equation}
Above we took the charges of the trigger field $\phi$ and  the inflaton $\sigma$ to be $-1$ and $0$ respectively. Note that the charge of $\sigma$ has to vanish but that of $\phi$ can be
different as long as it is the opposite of $TrQ_A$. (This requires the existence of at least another field charged under the anomalous $U(1)$ with zero VEV imposed by other F and D constraints.)
We see that the potential
is very similar to that of hybrid inflation and reduces to it for small $\phi$, i.e. $\phi<<M_P$\cite{HI}.
There are two minima for this potential: one at $\phi=0$ and $\sigma$ free and the other
at $\phi=M$ and $\sigma=0$. 

Hybrid inflation occurs for large $\sigma$, $\sigma \sim M<<M_P$ and small $\phi$, $\phi<<M$.
Then the vacuum energy is dominated by the anomalous D--term  
\begin{equation}
V_0 \sim g^2 M^4= 3H^2 M_P^2
\end{equation}
Thus, the Hubble constant is given by $H^2 \sim g^2 M^4/M_P^2$. In this regime
the masses for $\phi$ and $\sigma$  are easily found to be
\begin{equation}
m^2_{\phi}=4|c|^2 \sigma^2-2g^2 M^2 
\end{equation}
and
\begin{equation}
m^2_{\sigma}=4|c|^2 \phi^2
\end{equation}
We see that for large $\sigma$, i.e $\sigma^2>g^2M^2/2|c|^2=\sigma^2_{cr}$, $m^2_{\phi}$ is positive and
$m^2_{\phi}>>H^2$. As a result,  $\phi$ settles to its minimum at $\phi=0$ very quickly.
At the minimum,  the tree level $\sigma$ mass vanishes. However since supersymmetry
is broken by the D--term there is a mass splitting in the $\phi$ supermultiplet. At one loop
this leads to a potential for $\sigma$ given by\cite{DVA,DSS}
\begin{equation}
V_{1-loop}=g^2 M^4 \left(1+{g^2 \over {16\pi^2}} log {c^2 \sigma^2 \over \Lambda^2} \right)
\end{equation}
 $\Lambda$ is the renormalization scale which does not affect physical quantities.
The resulting $\sigma$ mass is
\begin{equation}
m^2_{\sigma}={g^4 M^4 \over 16 \pi^2 \sigma^2}< ({c^2 \over 8 \pi^2}) g^2 M^2 
\end{equation}

where the second inequality comes from $\sigma \leq \sigma_{cr}$. The slow--roll condition
$m^2_{\sigma}<<H^2$ is satisfied for $M^2>>|c|^2M_P^2/24 \pi^2$. When this holds
 $\sigma$ rolls down the potential very slowly during inflation as required. 
The slow--roll condition is can be satisfied since $m_{\sigma}$ arises at one--loop.
Inflation ends when
$m^2_{\sigma} \sim H^2$ which happens around $\sigma^2 \sim g^2 M_P^2/16 \pi^2$. Then
$\sigma$ begins to roll down the potential quickly. Once $\sigma^2<g^2M^2/2|c|^2$, $m^2_{\phi}$ becomes
negative and large compared to $H$.  $\phi=0$ becomes a maximum and $\phi$ begins to roll to its true
minimum at $\phi=M$. In the meantime $\sigma$ decreases to its true minimum at $\sigma=0$.
The vacuum energy vanishes at this minimum because the D--term vanishes and supersymmetry is restored; the loop corrections to the potential vanish in this case.
This is essentially the hybrid inflation scenario and we obtained it from an anomalous D-term
in supergravity. The superpotential and the F--terms were needed only to obtain the
$\phi^2 \sigma^2$ interaction which is crucial for hybrid inflation.

One must further check that the density fluctuations for $g \sim 1/2$ and parameters satisfying
the slow--roll condition  $M^2>>|c|^2M_P^2/24 \pi^2$ are of the correct magnitude\cite{HI}

\begin{equation}
{\delta \rho \over \rho} \sim {5 c M^5 \over {M_{Planck}^3 m^2_{\sigma}}} \sim 5 \times10^{-5}
\end{equation}
as required by COBE data. From Eq. (4) we see that $M \sim M_{Planck}/300$ for  
 $TrQ_A \sim 1$. Using Eq. (14) for  $m^2_{\sigma}$ at $\sigma=\sigma_{cr}$,
we find that we obtain the correct magnitude of density perturbations if $c \sim 1/10$. We
remind that $c$ is $g$ times a number of order of unity, i.e. $c \sim 1/2$ although it is certainly possible to have such small coefficients. It is encouraging to find that both the slow--roll
and ${\delta \rho / \rho}$ constraints can be satisfied since this is a nontrivial task in
a model with only two parameters, $M$ and $c$ highly restricted by its stringy origin.

\section{String Derived Supergravity Models}

The first requirement for the above scenario for hybrid inflation is the existence of an anomalous
$U(1)$ gauge group. There are large classes of string models which indeed have such 
anomalous $U(1)$s. This  field theoretical anomaly is cancelled by a stringy Green--Schwarz
mechanism. We also assumed that $M$ is fixed by $TrQ_A \sim 1$ which is much smaller
than the generic anomaly in realistic string models $TrQ_A \sim 100$. However, there is
no theoretical reason for dismissing the possibility of string models with such small anomalies.
 
We considered a specific supergravity toy model with the superpotential and Kahler
potential given by Eqs. (6) and (7). In this section, we argue that such a model can arise naturally
in supergravity models derived from free fermionic strings\cite{FFS}. The string derived superpotential
should only contain one term (given by Eq. (6)) for the two fields $\phi$ and $\sigma$. 
This can be achieved in principle by using discrete symmetries even though it is somewhat
unnatural in string models. The other and more important ingredient of our supergravity toy model is the Kahler potential given by Eq. (7). We will now see that some free fermionic string models can have this type of Kahler potential. 

The massless string spectrum can be divided into two parts; the untwisted and twisted sectors.
In realistic free fermionic models which are based on $Z_2 \times Z_2$ twists, the untwisted
sector contains three sectors; one related to each twist of  $Z_2 \times Z_2$. Each such sector 
can contain zero, one or two modulus--like fields depending on the boundary condition
vectors which define the string model\cite{EH}. (These fields are not necessarily string moduli so that they can appear in the perturbative superpotential.) For an untwisted sector which contains only one modulus
the Kahler potential is found to be\cite{NL}
\begin{equation}
K=-log(\sigma+ \bar \sigma) + K(\psi,\sigma)
\end{equation}
where $\sigma$ is modulus--like field and $\psi$ are matter fields in this sector which
do not concern us. The above Kahler potential is exactly what we need in our model.
If there are no modulus--like fields in a sector, $K$ is given by the term for the
matter fields, $K=\psi+ \bar \psi$.
The other term in  $K$ in Eq. (7) is a minimal Kahler potential for $\phi$. This can be obtained for the twisted sector matter fields with the Kahler potential\cite{NL}
\begin{equation}
K= \phi_1 \bar \phi_1 e^{(K_2+K_3)/2}+ cyclic \quad permutations
\end{equation}
Here the subscript denotes the sector, e.g. the Kahler potential of the first twisted sector 
depends on the Kahler potential of the second and third untwisted sectors etc. If there are two
untwisted sectors without moduli--like fields so that their Kahler potentials are minimal in
matter fields $K=\psi+ \bar \psi$,
we get a minimal Kahler potential for $\phi$, $K(\phi_1)= \phi_1 \bar \phi_1 e^{(\psi_2+ \bar \psi_2)(\psi_3+ \bar \psi_3)}$. Assuming that theese matter fields have vanishing VEVs during inflation we obtain a minimal Kahler potential for $\phi$.

Thus we find that the Kahler potential we assumed can be obtained in free fermionic string derived supergravity models which have only one modulus--like field and two sectors without
such fields. In such models $\sigma$ of our model is a modulus--like field whereas $\phi$ is a
twisted matter field. 

In string models, the possible terms in the cubic superpotential are severely constrained by
the world--sheet conformal field theory selection rules (in addition to discrete symmetries) so
that not many types of terms are allowed.
Amazingly, for free fermionic models of the above kind, among the few possible cubic terms 
there is one of the type $\sigma \phi^2$ (i.e. a term containing an untwited modulus--like field
with two twisted fields from the same sector)
as we assumed. We found that the coefficient of this term has to be somewhat small $c \sim 1/10$ in order to get the correct density perturbations.
Even though this is not a natural value for $c$ it is certainly not impossible.

\section{Discussion and Conclusions}

In this letter, we argued that the main obstacle for having inflation in supergravity namely
the inflaton mass problem can naturally be solved if the vacuum energy is dominated by a
D--term rather than an F--term.
We showed that hybrid inflation can occur in supergravity if it is driven 
 by an anomalous D--term associated with an anomalous
$U(1)$ gauge group. In this case, the mass of the inflaton can be much 
smaller than the Hubble constant and inflation can take place since the slow--roll
condition $m^2_{\sigma}<<H^2$ can be satisfied. If the inflaton is neutral under the anomalous $U(1)$, it has vanishing mass at tree level
and obtains a small (i.e. $<<H$) mass only at the one--loop level.
This is in contrast to the problematic case in which
the vacuum energy is dominated by the F--term contribution to the scalar potential and $m^2_{\sigma} \sim H^2$.
The anomalous D-term contribution gives rise to the vacuum energy and drives inflation. In addition, a simple cubic superpotential gives the $\phi^2 \sigma^2$ interaction that is essential for hybrid inflation. In order to obtain the correct magnitude of 
density perturbations, the
coefficient of this term and the coefficient of the anomaly must be somewhat small. Even though
these are not the natural values of these parameters, it is encouraging that all requirements
for inflation can be satisfied. We stress that this is highly nontrivial in a model with only two parameters $M$ and $c$ and which is constrained by its stringy origin.
 
A supergravity D--term can drive inflation only if there is anomalous contribution.
Such anomalous D--terms are generic to certain classes of string derived supergravity models.
We also showed that this kind of model (i.e. the Kahler potential and the superpotential) 
can arise in some realistic supergravity models derived from free fermionic strings. A complete check on a realistic string model is very diffcult to make due to the large number of fields and the complexity of the
superpotential. Nevertheless we expect the essential features of our toy model to survive
at least in some string models.

\section{Acknowledgements}
We would like to thank J. G. Bellido, A. Casas, M. Dine, D. Wands and especially A. Linde for
very useful discussions.

\end{document}